# Role of parity transformation for the fluctuation theorem: Limit cycle and symmetry breaking

Yasuyuki Matsuo

*Department of Physical Engineering, Mie University, Mie 514-8507, Japan*

**Abstract** The fluctuation theorem of the Crooks type is studied for thermodynamic nonlinear- multivariate systems. In particular, a bivariate system having a limit cycle is discussed in detail. It is explicitly shown how the time reversal operation has to be combined with the parity transformation in order to derive the fluctuation theorem for the change of the "renormalized entropy". Furthermore, breakdown of the symmetry in a limit-cycle model due to the effect of fluctuations is also discussed.

## I. INTRODUCTION

Ever since the discovery of the fluctuation theorems [1,2], a number of discussions have been developed about their various theoretical aspects as well as applications (for a recent review article, see Ref. [3]). Among others, the work in Ref. [4] is of interest for us, since it is based on the Onsager-Machlup theory of fluctuations [5], where macroscopic (i.e., thermodynamic) quantities are treated and the linearity of relaxation processes is assumed. The authors of Ref. [4] introduce an external dragging into the Onsager-Machlup theory in order to realize a stationary state and rederive the theorems for fluctuations of work, friction, and quantity of heat. However, the fundamental importance of the theorems is concerned with states far from equilibrium, where the linear approximation is not valid any more, since the stationary state may be realized by the dynamics itself. This point has recently been discussed in Ref. [6]. There, it has been found through a nonlinear generalization of the Onsager-Machlup theory that a quantity satisfying a fluctuation theorem of the Crooks type [7] is not for the ordinary entropy change but for the "renormalized" one. This opens a possibility of understanding the notion of entropy in the strongly nonequilibrium regime and suggests that consideration of the nonlinear dynamics can further reveal the physics of fluctuations.

In this paper, we study the fluctuation theorem for the change of the renormalized entropy in a system described by macroscopic thermodynamic multivariables. In particular, we do so by employing, as an analytically tractable example, a nonlinear system having a limit cycle, which represents a nonequilibrium stable state. We explicitly show how it is necessary to combine the time reversal operation with the parity transformation in order to derive the fluctuation theorem. In addition, we find that the circular symmetry of the limit cycle becomes broken due to the effect of fluctuations.

The paper is organized as follows. In Sec. II, we generalize the work of Ref. [6] to a general nonlinear multivariate case. In Sec. III, we study an explicit example of a limit cycle in a bivariate system and show how the parity transformation plays an important role in deriving the fluctuation theorem for the change of the renormalized entropy. Then, we analyze the physical property of the most probable process in terms of the thermodynamic Lagrangian and show that the circular symmetry of the limit cycle can be broken due to the noise effect. Section IV is devoted to concluding remarks.

## II. THERMODYNAMIC LAGRANGIAN OF MULTIVARIATE SYSTEMS

Let us consider a system, which is characterized by a set of thermodynamic quantities, $\phi_i (i=1,2,...,n)$ (e.g., internal energy, volume, numbers of particles, etc.) and is surrounded by the environment. These quantities evolve in time according to certain dynamics, which describes a process of relaxation to a *stationary state* of the system. Such dynamics may be represented by the following coupled Langevin equations:

$$\frac{d\phi_i}{dt} = F_i(\phi) + \xi_i. \qquad (1)$$

Here $F_i(\phi)$'s are the elements of the current vector and $\xi_i$'s are noise terms describing the environmental effect. For the sake of simplicity, we assume $\xi_i$'s to be the Gaussian white noises satisfying

$$\overline{\xi_i(t)} = 0, \quad \overline{\xi_i(t)\xi_j(t')} = 2D_{ij}\delta(t-t'), \qquad (2)$$

where the overbar stands for the average over the Gaussian-noise distribution, and $D_{ij}$ is a regular symmetric diffusion-constant matrix.

The dynamics given by Eq. (1) is a generalization of the work in Ref. [6] as well as the Onsager-Machlup theory [5]. The Fokker-Planck equation associated with Eq. (1) reads

$$\frac{\partial \rho(\phi,t)}{\partial t} = -\sum_{i=1}^{n} \frac{\partial}{\partial \phi_i} \left[ F_i(\phi) \rho(\phi,t) - \sum_{j=1}^{n} D_{ij} \frac{\partial \rho(\phi,t)}{\partial \phi_j} \right], \quad (3)$$

where $\phi \equiv (\phi_1, \phi_2, \cdots, \phi_n)$ and $\rho(\phi,t) d^n \phi$ is the probability of finding the system in the "volume", $[\phi_1, \phi_1 + d\phi_1] \times [\phi_2, \phi_2 + d\phi_2] \times \cdots \times [\phi_n, \phi_n + d\phi_n]$, at time $t$. We take a time interval, $[0, \tau]$, and impose the temporal boundary conditions, $\phi(0) = X$ and $\phi(\tau) = Y$. The forward transition probability from $X$ to $Y$ is given by the following functional integral:

$$f_F(Y, \tau | X, 0) = N \int D\xi \int_{\phi(0)=X}^{\phi(\tau)=Y} D\phi \, \delta[\phi - \phi_\xi] \exp\left( -\frac{1}{4} \sum_{i,j=1}^{n} \int dt \, D_{ij}^{-1} \xi_i(t) \xi_j(t') \right)$$

$$= N \int_{\phi(0)=X}^{\phi(\tau)=Y} D\phi \, \text{Det}\left[ \left( \delta_{ij} \frac{d}{dt} - \frac{\partial F_i(\phi)}{\partial \phi_j} \right) \delta(t-t') \right]$$

$$\times \exp\left\{ -\frac{1}{4} \sum_{i,j=1}^{n} \int_0^\tau dt \, D_{ij}^{-1} \left[ \frac{d\phi_i}{dt} - F_i(\phi) \right] \left[ \frac{d\phi_j}{dt} - F_j(\phi) \right] \right\}, \quad (4)$$

where the subscript "$F$" denotes the forward process, $N$ is a normalization factor, $D_{ij}^{-1}$ the inverse of the diffusion-constant matrix, $\phi_{i\xi}(t)$ the solution of Eq. (1), and $\delta[\phi - \phi_\xi] \equiv \prod_{i,t} \delta(\phi_i(t) - \phi_{i\xi}(t))$. To evaluate the functional determinant, we employ the formula, $\text{Det } M = \exp(\text{Tr} \ln M)$, with

$$M_{ij}(t, t') \equiv \left[ \delta_{ij} \frac{d}{dt} - \frac{\partial F_i(\phi)}{\partial \phi_j} \delta(t-t') \right]. \tag{5}$$

We rewrite $M_{ij}(t, t')$ in the form: $M_{ij}(t, t') \equiv (d/dt) K_{ij}(t, t')$, where

$$K_{ij}(t, t') = \delta_{ij} \delta(t-t') + \frac{\partial F_i(\phi(t'))}{\partial \phi_j(t')} G(t-t') \tag{6}$$

with $G(t) = \int_{-\infty}^{\infty} d\omega \widetilde{G}(\omega) e^{i\omega t}$ being Green's function satisfying $dG(t)/dt = \delta(t)$. Here, imposing the condition that $\widetilde{G}(\omega)$ is analytic in the complex-$\omega$ upper half plane as in the dispersion relation, we have

$$G(t-t') = -\theta(t'-t), \tag{7}$$

where $\theta(t)$ is the Heaviside step function. Absorbing $\exp[\mathrm{Tr}\ln(d/dt)]$ in the normalization factor and expanding the logarithm, we have

$$\ln(\mathrm{Det}\, K) = \theta(0) \sum_{i=1}^{n} \int_0^{\tau} dt \frac{\partial F_i(\phi(t))}{\partial \phi_i(t)} - \frac{1}{2} \sum_{i,j=1}^{n} \int_0^{\tau} dt_1 \int_0^{\tau} dt_2 \theta(t_1-t_2) \theta(t_2-t_1)$$

$$\times \frac{\partial F_i(\phi(t_1))}{\partial \phi_i(t_1)} \frac{\partial F_j(\phi(t_2))}{\partial \phi_j(t_2)} + \cdots, \tag{8}$$

In this series, only the first term survives because the integrals including the products of the step functions vanish. Setting $\theta(0) = 1/2$, we obtain

$$\mathrm{Det}\, M \propto \exp\left( \frac{1}{2} \sum_{i=1}^{n} \int_0^{\tau} dt \frac{\partial F_i(\phi)}{\partial \phi_i} \right), \tag{9}$$

Consequently, we have the following form for the forward transition probability:

$$f_F(Y,\tau|X,0) = N \int_{\phi(0)=X}^{\phi(\tau)=Y} D\phi \exp\left(-\int_0^\tau dt\, L\right), \qquad (10)$$

where

$$L = -\frac{1}{4}\sum_{i,j=1}^n \left\{D_{ij}^{-1}\left[\frac{d\phi_i}{dt} - F_i(\phi)\right]\left[\frac{d\phi_j}{dt} - F_j(\phi)\right]\right\} - \frac{1}{2}\sum_{i=1}^n \frac{\partial F_i(\phi)}{\partial \phi_i}. \qquad (11)$$

is the quantity referred to as the thermodynamic Lagrangian.

## III. PARITY TRANSFORMATION, FLUCTUATION THEOREM, AND SYMMETRY BREAKING: A LIMIT CYCLE

In this section, we apply the general discussion developed in the preceding section to the case of a bivariate system having a limit cycle and study the fluctuation theorem. We also discuss how the geometric structure of the limit cycle can be deformed by the noises.

### A. Fluctuation theorem

Limit cycles are nonequilibrium stable solutions of multivariate coupled nonlinear equations. They can be found in important systems and phenomena including oscillating chemical reactions, predator-prey ecological system, and nerve activity [8]. An oscillating chemical reaction (more specifically, the Belousov-Zhabotinsky reaction) and the relevant fluctuation theorem have been studied in Refs. [9,10], but these works are not directly related to our discussion below. In our case, the basic variables are of thermodynamics. Therefore, a limit cycle we are going to consider may be analogous to a thermodynamic engine experiencing an endoreversible process in nonequilibrium thermodynamics [11,12]. Our main purposes of discussing a limit cycle are (i) to see how the standard fluctuation theorem has to be modified due to nonlinearity and (ii) to clarify the physical properties of fluctuations in a stationary state far from equilibrium.

Let us find a stationary solution of Eq. (3) in the bivariate case. For this purpose, it is convenient to use the Helmholtz-Hodge method to decompose $F_i$ into the gradient part and the divergence-free part:

$$F_i(\phi) = \sum_{j=1}^{2} D_{ij} \frac{\partial \Sigma}{\partial \phi_j} + \Gamma_i, \tag{12}$$

$$\sum_{i=1}^{2} \frac{\partial \Gamma_i}{\partial \phi_i} = 0. \tag{13}$$

Equation (13) is automatically fulfilled, if we set

$$\Gamma_i = \sum_{j=1}^{2} \varepsilon_{ij} \frac{\partial \Lambda}{\partial \phi_j}, \tag{14}$$

where $\varepsilon_{ij}$ is the Levi-Civita symbol: $\varepsilon_{ij} = -\varepsilon_{ji}$ and $\varepsilon_{12} = 1$. We look for a stationary solution of the form,

$$\rho_S(\phi) \propto e^{\Sigma(\phi)}. \tag{15}$$

Later, we will see that the quantity, $\Sigma$, plays a role of the renormalized entropy proposed in Ref. [6]. Substituting Eqs. (12) and (14) into Eq. (3) with $n = 2$, we find that the stationarity condition, $\partial \rho / \partial t = 0$, yields

$$\sum_{i,j=1}^{2} \varepsilon_{ij} \frac{\partial \Lambda}{\partial \phi_i} \frac{\partial \Sigma}{\partial \phi_j} = 0. \tag{16}$$

This implies that both $\Sigma$ and $\Lambda$ depend on $\phi_1$ and $\phi_2$ through a single differentiable function, $r = r(\phi_1, \phi_2)$.

Now, let us apply the above discussion to a system having a limit cycle. Here, as an analytically tractable example, we study the following coupled nonlinear differential equations subject to the noises

$$\begin{cases} \dfrac{d\phi_1}{dt} = \phi_2 + \phi_1(1-\phi_1^2-\phi_2^2) + \xi_1 \\ \dfrac{d\phi_2}{dt} = -\phi_1 + \phi_2(1-\phi_1^2-\phi_2^2) + \xi_2 \end{cases}, \quad (17)$$

where the variables are made dimensionless. Clearly, the elements of the current vector are $F_1(\phi) = \phi_2 + \phi_1(1-\phi_1^2-\phi_2^2)$ and $F_2(\phi) = -\phi_1 + \phi_2(1-\phi_1^2-\phi_2^2)$. If the noises are switched off, then this system has a limit cycle defined by $\phi_1^2 + \phi_2^2 = 1$, which possesses an obvious circular symmetry. For the sake of simplicity, we assume that the diffusion-constant matrix has the form: $D_{ij} = D\delta_{ij}$. Then, the Fokker-Planck equation in Eq. (3) becomes

$$\frac{\partial \rho(\phi,t)}{\partial t} = -\sum_{i=1}^{2} \frac{\partial}{\partial \phi_i}\left[ F_i(\phi)\rho(\phi,t) - D\frac{\partial \rho(\phi,t)}{\partial \phi_i} \right]. \quad (18)$$

And, from Eq. (17), $\Sigma$ and $\Lambda$ are found to be

$$\Sigma(\phi) = \frac{r^2}{2D} - \frac{r^4}{4D}, \quad (19)$$

$$\Lambda(\phi) = \frac{r^2}{2}, \quad (20)$$

where

$$r \equiv \sqrt{\phi_1^2 + \phi_2^2}, \quad (21)$$

in the present case. Equation (16) is automatically fulfilled, since $\Sigma = \Sigma(r)$ and $\Lambda = \Lambda(r)$, and the stationary solution of the form in Eq. (15) is obtained.

The thermodynamic Lagrangian associated with Eq. (17) is given by

$$L = \sum_{i=1}^{2}\left[\frac{1}{4D}\left(\frac{d\phi_i}{dt} - F_i(\phi)\right)^2 - \frac{1}{2}\frac{\partial F_i(\phi)}{\partial \phi_i}\right].\qquad(22)$$

In what follows, we are going to prove the fluctuation theorem. As we shall see, not only the time reversal but also the parity transformation plays a crucial role.

Consider the time reversal (with the shift): $t \to \hat{t} = \tau - t$. Under this operation, the thermodynamic Lagrangian in Eq. (22) transforms as

$$L\left(\phi(t), \frac{d\phi(t)}{dt}\right) = L\left(\hat{\phi}(\hat{t}), \frac{d\hat{\phi}(\hat{t})}{d\hat{t}}\right)$$
$$+ \frac{1}{D}\sum_{i=1}^{2}\left[\frac{d\hat{\phi}_i(\hat{t})}{d\hat{t}}\left(D\frac{\partial \Sigma(\hat{\phi}(\hat{t}))}{\partial \hat{\phi}_i(\hat{t})} + \sum_{j=1}^{2}\varepsilon_{ij}\frac{\partial \Lambda(\hat{\phi}(\hat{t}))}{\partial \hat{\phi}_j(\hat{t})}\right)\right],\qquad(23)$$

provided that $\phi$ is assumed to be scalar: $\phi(t) = \hat{\phi}(\hat{t})$. We note that the second term on the right-hand side still does not have the form of a total derivative in time, implying that *the fluctuation theorem cannot be derived as long as only the time reversal is employed* (compare this situation with that in Ref. [6]).

However, this difficulty can be overcome if we combine the time reversal with the parity transformation in two dimensions:

$$(\hat{\phi}_1, \hat{\phi}_2) \to (\bar{\phi}_1, \bar{\phi}_2) \equiv (\hat{\phi}_1, -\hat{\phi}_2).\qquad(24)$$

Under this operation combined with the above-mentioned time reversal, the thermodynamic Lagrangian is found to transform as

$$L\left(\phi(t), \frac{d\phi(t)}{dt}\right) = L\left(\bar{\phi}(\hat{t}), \frac{d\bar{\phi}(\hat{t})}{d\hat{t}}\right) + \frac{d\Sigma(\bar{\phi}(\hat{t}))}{d\hat{t}}.\qquad(25)$$

Therefore, the change is a total derivative in time, as desired. Substituting Eq. (25) into Eq. (10), we obtain the detailed balance condition

$$f_F(Y,\tau|X,0)\rho_S(X) = f_F(X,\tau|Y,0)\rho_S(Y), \qquad (26)$$

where $\rho_S$ is given in Eq. (15), and the temporal boundary conditions are $X = (\bar{\phi}_1(\tau), \bar{\phi}_2(\tau))$ and $Y = (\bar{\phi}_1(0), \bar{\phi}_2(0))$.

Finally, we consider the reverse transition probability. Usually, such a quantity is a transition probability, in which the time derivative of $\phi$ is inverted and the initial state is replaced by the final one [7]. However, the above discussion forces us to also include the parity transformation. Thus, the reverse transition probability should be defined by

$$f_R(X,\tau|Y,0) = N \int_{\bar{\phi}(\tau)=X}^{\bar{\phi}(0)=Y} D\bar{\phi} \exp\left(-\int_0^\tau d\hat{t}\, \tilde{L}\right)$$

$$= f_F(Y,\tau|X,0), \qquad (27)$$

where $\tilde{L}$ is the transformed thermodynamic Lagrangian given by

$$\tilde{L} = \sum_{i=1}^{2}\left[\frac{1}{4D}\left(\frac{d\bar{\phi}_i}{d\hat{t}} + D\frac{\partial \Sigma}{\partial \bar{\phi}_i} - \sum_{j=1}^{2} \varepsilon_{ij}\frac{\partial \Lambda}{\partial \bar{\phi}_j}\right)^2 - \frac{1}{2}\frac{\partial F_i(\bar{\phi})}{\partial \bar{\phi}_i}\right]. \qquad (28)$$

Upon deriving Eq. (27), we have used $\hat{t} = \tau - t$ with $0 \leq t \leq \tau$.

With these preparations, it is now straightforward to prove the fluctuation theorem of the Crooks type. The probability that the change of $\Sigma$ during the time interval $[0,\tau]$ is $\Delta\Sigma$ is calculated to be

$$P_F(\Delta\Sigma) = \left\langle \delta\left(\Delta\Sigma - \int_0^\tau dt\, \frac{d\Sigma(\phi)}{dt}\right)\right\rangle_F$$

$$\equiv \iint dXdY\, \delta(\Delta\Sigma - [\Sigma(Y) - \Sigma(X)])\, f_F(Y,\tau|X,0)\rho(X,0). \tag{29}$$

Using Eqs. (26) and (27), we obtain the theorem:

$$\frac{P_F(\Delta\Sigma)}{P_R(-\Delta\Sigma)} = e^{\Delta\Sigma}, \tag{30}$$

where $P_R(-\Delta\Sigma)$ is defined by [7]

$$P_R(-\Delta\Sigma) \equiv \iint dXdY\, \delta(-\Delta\Sigma - [\Sigma(Y) - \Sigma(X)])\, f_R(X,\tau|Y,0)\rho(Y,0) \tag{31}$$

This is the first result of the present work. It generalizes the work in Ref. [6] in such a way that importance of the parity transformation is explored and the renormalized entropy, $\Sigma$, is identified in a multi-dimensional system based on the Helmholtz-Hodge decomposition.

### B. Noise-induced symmetry breaking

To evaluate the effects of the noises on the circular symmetry of the limit cycle, we analyze the "most probable process" [4,5], which is given by the solutions of the Euler-Lagrange equations obtained from the thermodynamic Lagrangian. It is convenient to rewrite Eq. (22) in the following form:

$$L = \frac{1}{4D}\left[\left(\frac{dr}{dt}\right)^2 + r^2\left(\frac{d\theta}{dt} + 1\right)^2 + r^2(1-r^2)^2\right] - \frac{1}{2}\frac{d\Sigma}{dt} - 1 + 2r^2, \tag{32}$$

where $r$ is given in Eq. (21) and $\theta = \tan^{-1}(\phi_2/\phi_1)$, which is a cyclic coordinate. The equations for $r$ and $\theta$ are

$$\frac{d^2r}{dt^2} - (1+8D)r + 4r^3 - 3r^5 - \frac{4D^2l^2}{r^3} = 0, \tag{33}$$

$$\frac{r^2}{2D}\left(\frac{d\theta}{dt}+1\right)=l,\qquad(34)$$

respectively, where $l$ is a constant. From these, we obtain the following "energy integral":

$$\frac{1}{2}\left(\frac{dr}{dt}\right)^2-\frac{1+8D}{2}r^2+r^4-\frac{1}{2}r^6+\frac{2D^2l^2}{r^2}=\alpha,\qquad(35)$$

where $\alpha$ is also a constant. Introducing the variable, $R=r^2$, we rewrite Eq. (35) as follows:

$$\left(\frac{dR}{dt}\right)^2-8\alpha R-4(1+8D)R^2+8R^3-4R^4=-16D^2l^2.\qquad(36)$$

This equation can also be regarded as the energy integral with the quartic potential:

$$V(R)=-8\alpha R-4(1+8D)R^2+8R^3-4R^4.\qquad(37)$$

There exist two distinct cases depending on the sign of the discriminant of $dV/dR$:

$$\Delta_{dV/dR}=-108\alpha^2-864\alpha D+(1+8D)^2(1-64D).\qquad(38)$$

If $\Delta_{dV/dR}<0$, then $V$ is a single barrier potential, whereas it is a double barrier potential if $\Delta_{dV/dR}>0$. The former is a trivial case, where the one and only nondivergent solution in the limit $t\to\infty$ is the one that relaxes to $R=\text{const}$, i.e., a class of limit cycles with the circular symmetry. The latter is the nontrivial one, in which we are interested. In this case, oscillation can occur between the two barriers (see Ref. [13] for an analytical discussion), and therefore the circular symmetry is broken. The border, $\Delta_{dV/dR}=0$, has two solutions. In Fig. 1, we present the plots of them in the $D\alpha$ plane. The symmetry breaking is realized in the "triangular" region. In Fig. 2, we present an example of the oscillation breaking the circular symmetry. Since the renormalized entropy, $\Sigma$, is a function of $r$ (thus, $R$), its value also oscillates. This is our second result.

## IV. CONCLUDING REMARKS

We have studied the fluctuation theorem of the Crooks type for the change of the renormalized entropy in nonlinear multivariate systems by generalizing the work in Ref. [6]. Employing a bivariate system having a limit cycle as an explicit example, we have shown how the parity transformation should be combined with the time reversal operation in order for the theorem to be proved. We have also analytically studied in that system the circular symmetry of the limit cycle can be broken due to the noise effects from the environment.

The necessity of combining the time reversal with the partial inversion of the microscopic coordinate variables, which is analogous to the parity transformation in Eq. (24), for deriving a steady-state fluctuation theorem has been discussed in the works in Ref. [1], where the shear stress of a fluid with the preferred anisotropy is considered. In those works, such a partial inversion is indivisibly connected to this anisotropy. On the other hand, in our case, the variables are macroscopic thermodynamic ones, and the limit cycle has no preferred anisotropy. In addition, what we have studied here is not a steady-state fluctuation theorem but a fluctuation theorem of the Crooks type. These facts seem to suggest the universal necessity of combining the time reversal with the parity operation or inversion. This reminds us of the CPT invariance of the fundamental physical laws, although its relevance to nonequilibrium statistical mechanics is not known at all yet. Clearly, further investigations are needed to clarify whether the present transformation property is a remnant of such invariance.

Since we have treated the macroscopic thermodynamic variables, as mentioned in Sec. III, limit cycles can be interpreted as endoreversible processes in nonequilibrium thermodynamics. We expect that the present work may contribute to elucidating the significance of fluctuations for such processes, which are supposed to play an important role in small systems.

## ACKNOWLEDGMENTS

The author would like to thank Professor Sumiyoshi Abe for helpful discussions and comments. He also thanks the anonymous referee for drawing his attention to the works in Ref. [1], where the role of the partial inversion is discussed for microscopic phase-space variables.

---

Figure Caption

FIG. 1. Plots of the solutions of $\Delta_{dV/dR} = 0$. There are two solutions, the solid and dashed curves. The oscillation is realized in the region surrounded by these curves and the vertical axis.

FIG. 2. Plot of the oscillatory solution corresponding to the values of the parameters, $\alpha = -0.05$, $D = 0.01$, and $l = 0$. All quantities are dimensionless.

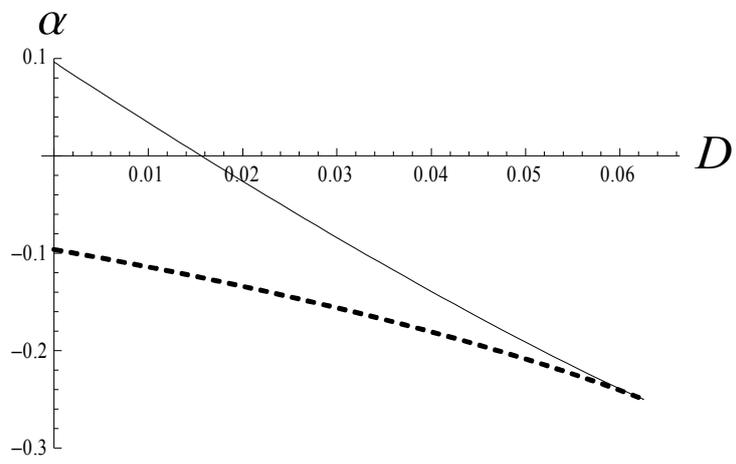

FIG. 1

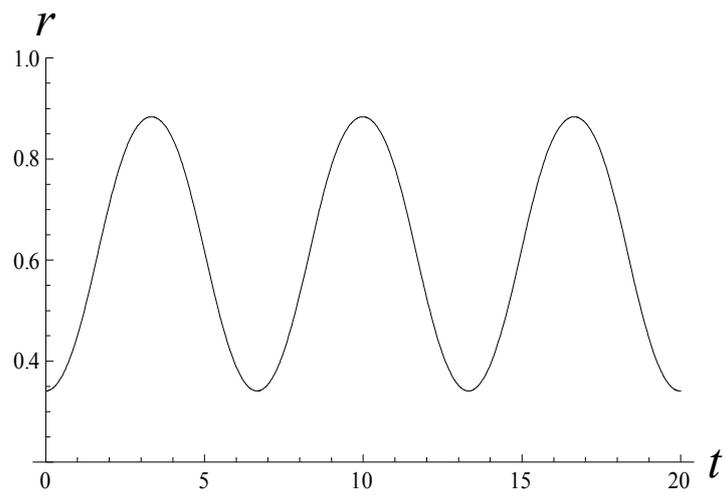

FIG. 2